\documentclass[a4paper]{article}
\usepackage{listings}
\usepackage{xcolor}
\usepackage{multirow}
\usepackage{jheppub}  
\usepackage{dcolumn}
\usepackage[english]{babel}
\usepackage[utf8x]{inputenc}
\usepackage[T1]{fontenc}
\usepackage{palatino}
\usepackage{amsmath}
\usepackage{afterpage}
\usepackage{graphicx, subcaption}
\usepackage[colorinlistoftodos]{todonotes}
\usepackage[colorlinks=true]{hyperref}
\usepackage{makeidx}
\newcommand{\be}{\begin{equation}}  
\newcommand{\ee}{\end{equation}}  
\newcommand{\bea}{\begin{eqnarray}}  
\newcommand{\eea}{\end{eqnarray}} 
\newcommand{\drv}{{\rm d}}

\newcommand{\tcite}[1]{~\cite{#1}}
\newcommand{\tref}[1]{~\ref{#1}}

\makeindex

\begin{document}

\vspace*{1.2cm}

\thispagestyle{empty}
\begin{center}
{\LARGE \bf Phenomenology of the
hadronic structure at small-$x$}

\par\vspace*{7mm}\par

{

\bigskip

\large \bf Francesco Giovanni Celiberto}

\bigskip

{\large \bf  E-Mail: fceliberto@ectstar.eu}

\bigskip

{
European Centre for Theoretical Studies in Nuclear Physics and Related Areas (ECT*),
\\
I-38123 Villazzano, Trento, Italy
\vskip .18cm
Fondazione Bruno Kessler (FBK)
I-38123 Povo, Trento, Italy
\vskip .18cm
INFN-TIFPA Trento Institute of Fundamental Physics and Applications
\\
I-38123 Povo, Trento, Italy
}

\bigskip

{\it Presented at the Low-$x$ Workshop, Elba Island, Italy, September 27--October 1 2021}

\vspace*{15mm}

\end{center}
\vspace*{1mm}

\begin{abstract}
:We present exploratory studies of the proton structure via two distinct kinds of gluon densities: the transverse-momentum dependent functions, whose evolution is determined by the CSS equation, and the unintegrated gluon distribution, whose definition is given in the BFKL formalism.
Our analyses are relevant for the exploration of the intrinsic motion of gluons inside protons in the small-$x$ regime, and for a comprehensive tomographic imaging of the proton at new-generation colliding facilities.
\end{abstract}

\vspace{-0.20cm}
\section{Introduction}
\label{sec:intro}
\vspace{-0.20cm}

Unraveling the inner structure of hadrons through a multi-dimensional study of their constituents represents a frontier research of phenomenological studies at new-generation colliding facilities.
The well-established collinear factorization that relies on a one-dimensional description of the proton content via collinear parton distribution functions (PDFs) has collected a long chain of achievements in describing data at hadronic and lepton-hadronic accelerators.

There are however vital questions on the dynamics of strong interactions which still do not have an answer.
Unveiling the origin of spin and mass of the nucleons requires a stretch of our vision from the collinear description to a tomographic viewpoint in three dimensions, naturally afforded by the \emph{transverse-momentum-dependent} (TMD) factorization.

In the small-$x$ regime a purely TMD-based approach may be not adequate, since large contributions proportional to $\ln (1/x)$ enter the perturbative series with a power that grows with the order, and need to be accounted for via an all-order resummation procedure.
The most powerful tool to resum those large logarithms is the Balitsky--Fadin--Kuraev--Lipatov (BFKL) formalism~\cite{Fadin:1975cb,Kuraev:1976ge,Kuraev:1977fs,Balitsky:1978ic} in the leading approximation (LL$x$), which means inclusion of all terms proportional to $\alpha_s^n \ln (1/x)^n$, and in the next-to-leading approximation (NLL$x$), including all terms proportional to $\alpha_s^{n+1} \ln (1/x)^n$.

In this paper we report progresses on the study of the proton structure at small-$x$ via two distinct kinds of gluon distributions.
In Section~\ref{sec:TMDs} we present the main features of a quite recent calculation of all (un)polarized gluon TMD distributions at leading twist, whose definition genuinely embodies BFKL effects.
In Section~\ref{sec:UGD} we provide with an evidence that helicity amplitudes for the exclusive leptoproduction of $\rho$-mesons act as discriminators for the existing models of the BFKL \emph{unintegrated gluon distribution} (UGD).

We come out with the message that forthcoming analyses at new-generation colliders will help to shed light on the hadronic structure at small-$x$ and to explore the interplay between different formalisms, such as the TMD factorization and the BFKL resummation.

\vspace{-0.20cm}
\section{Small-$x$ improved transverse-momentum dependent gluon distributions}
\label{sec:TMDs}
\vspace{-0.20cm}

The complete list of unpolarized and polarized gluon TMDs at leading twist (twist-2) was given for the first time in Ref.~\cite{Mulders:2000sh}. In Tab.~\ref{tab:gluon_TMDs} we present the eight twist-2 gluon TMDs for a spin-1/2 target, using the nomenclature convention as in Refs.~\cite{Meissner:2007rx,Lorce:2013pza}.
The two functions on the diagonal in Tab.~\ref{tab:gluon_TMDs} respectively represent the density of unpolarized gluons inside an unpolarized nucleon, $f_1^g$, and of circularly polarized gluons inside a longitudinally polarized nucleon, $g_1^g$.
In the collinear regime they correspond to the well-known unpolarized and helicity gluon PDFs.

TMD distributions receive contributions from the resummation of transverse-momentum logarithms which enter perturbative calculations. Much is know about this resummation~\cite{Bozzi:2003jy,Catani:2010pd,Echevarria:2015uaa}, but very little is known about the genuinely non-perturbative TMD content.
The distribution of linearly polarized gluons in an unpolarized hadron, $h_1^{\perp g}$, is particularly relevant, since it gives rise to spin effects in collisions of unpolarized hadrons~\cite{Boer:2010zf,Sun:2011iw,Boer:2011kf,Pisano:2013cya,Dunnen:2014eta,Lansberg:2017tlc}, whose size is expected to increase at small-$x$ values.
The Sivers function, $f_{1T}^{\perp g}$, gives us information about the density of unpolarized gluons in a transversely polarized nucleon, and plays a key role in the description of transverse-spin asymmetries that can be studies in collisions with polarized-proton beams.
Notably, in Ref.~\cite{Boussarie:2019vmk} it was argued that in the forward limit the Sivers function can be accessed in unpolarized electron-nucleon collisions via its connection with the QCD Odderon.

At variance with collinear distributions, TMDs are process-dependent via the \emph{gauge links} (or \emph{Wilson lines})~\cite{Brodsky:2002cx,Collins:2002kn,Ji:2002aa}.
Quark TMDs depend on the $[+]$ and $[-]$ staple links, which set the direction of future- and past-pointing Wilson lines, respectively. 
Gluon TMDs exhibit a more involved gauge-link dependence, since they are sensitive on combinations of two or more staple links. This brings to a more diversified kind of \emph{modified universality}.

Two major gluon gauge links appear: the $f\text{-type}$ and the $d\text{-type}$ ones. They are also known in the context of small-$x$ studies as Weisz\"acker--Williams and dipole structures, respectively.
The antisymmetric $f_{abc}$ QCD color structure appears in the $f$-type $T$-odd gluon-TMD correlator, whereas the symmetric $d_{abc}$ structure characterizes the $d$-type $T$-odd one. This fact leads to a dependence of $f$-type gluon TMDs on the $[\pm,\pm]$ gauge-link combinations, while $d$-type gluon TMDs are characterized by the $[\pm,\mp]$ ones.
Much more complicate, box-loop gauge links emerge in processes where multiple color exchanges connect both initial and final states~\cite{Bomhof:2006dp}. This leads to a violation of the TMD factorization~\cite{Rogers:2010dm} (see also Ref.\tcite{Rogers:2013zha}).

{
\renewcommand{\arraystretch}{1.7}

 \begin{table}
\centering
 \hspace{1cm} gluon pol. \\ \vspace{0.1cm}
 \rotatebox{90}{\hspace{-1cm} nucleon pol.} \hspace{0.1cm}
 \begin{tabular}[c]{|m{0.5cm}|c|c|c|}
 \hline
 & $U$ & circular & linear \\
 \hline
 $U$ & $f_{1}^{g}$ & & \textcolor{blue}{$h_{1}^{\perp g}$} \\
 \hline	
 $L$ & & $g_{1}^{g}$ & \textcolor{red}{$h_{1L}^{\perp g}$} \\
 \hline	
 $T$ & \textcolor{red}{$f_{1T}^{\perp g}$} & \textcolor{blue}{$g_{1T}^{g}$} & \textcolor{red}{$h_{1}^{g}$}, \textcolor{red}{$h_{1T}^{\perp g}$} \\
 \hline
  \end{tabular}
 \caption{A table of leading-twist gluon TMDs for spin-$1/2$ targets. 
 $U$, $L$, $T$ stand for unpolarized, longitudinally polarized and transversely polarized hadrons, whereas
 $U$, `circular', `linear' depict unpolarized, circularly polarized and linearly polarized gluons, respectively. 
 $T$-even (odd) functions are given in blue (red). 
 Black functions are $T$-even and survive the integration over the gluon transverse momentum.}
 \label{tab:gluon_TMDs}
 \end{table}

}

From the experimental point of view, the gluon-TMD sector is a largely unexplored field. 
First attempts at phenomenological analyses of the unpolarized gluon function have been presented in Refs.~\cite{Lansberg:2017dzg,Gutierrez-Reyes:2019rug,Scarpa:2019fol}. Phenomenological studies of the gluon Sivers TMD can be found in Refs.~\cite{Adolph:2017pgv, DAlesio:2017rzj,DAlesio:2018rnv,DAlesio:2019qpk}.
Therefore, exploratory analyses of gluon TMDs via simple and flexible models are needed. Pioneering studies along this direction were carried out in the so-called \emph{spectator-model} framework~\cite{Lu:2016vqu,Mulders:2000sh,Pereira-Resina-Rodrigues:2001eda}.
Formerly employed in the description of quark TMD distributions~\cite{Bacchetta:2008af,Bacchetta:2010si,Gamberg:2005ip,Gamberg:2007wm,Jakob:1997wg,Meissner:2007rx}, it is based on the assumption that the struck hadron with mass $\cal M$ and momentum $\cal P$ emits a gluon with longitudinal fraction $x$, momentum $p$, and transverse momentum $\boldsymbol{p}_T$, and what remains is treated as an effective on-shell particle having ${\cal M}_X$ and spin-1/2.
Within this model taken at tree level all the leading-twist TMDs in Table~\ref{tab:gluon_TMDs} can be calculated. Spectator-model gluon $T$-even densities were recently calculated in Ref.~\cite{Bacchetta:2020vty} and presented in Refs.~\cite{Celiberto:2021zww,Bacchetta:2021oht}.
In those works the nucleon-gluon-spectator vertex was modeled as follows
\begin{equation}
 \label{eq:form_factor}
 {\cal G}^{\, \mu} = \left( \tau_1(p^2) \, \gamma^{\, \mu} + \tau_2(p^2) \, \frac{i}{2{\cal M}} \sigma^{\, \mu\nu}p_\nu \right) \,,
\end{equation}
the $\tau_1$ and $\tau_2$ functions being dipolar form factors in $\boldsymbol{p}_T^2$. A dipolar profile for the couplings is useful to fade gluon-propagator divergences, quench large-$\boldsymbol{p}_T$ effects which are beyond the reach of a pure TMD description, and remove logarithmic singularities coming from $\boldsymbol{p}_T$-integrated distributions.
Furthermore, the spectator mass was allowed to take a continuous range of values weighed by a spectral function ${\cal S}_{\rm } ({\cal M}_X)$, which provides the necessary flexibility to reproduce both the small- and the moderate-$x$ shape of gluon collinear PDFs.
The analytic expression of the spectral function contains seven parameters and reads
\begin{equation}
\label{eq:rhoX}
 {\cal S}_{\rm } ({\cal M}_X) = \mu^{2a} \left( \frac{A}{B + \mu^{2b}} + \frac{C}{\pi \sigma} e^{-\frac{({\cal M}_X - D)^2}{\sigma^2}} \right) \,.
\end{equation}
Model parameters were fixed by performing a simultaneous fit of our unpolarized and helicity TMDs, $f_1^g$ and $g_1^g$, to the corresponding collinear PDFs from {\tt NNPDF}\tcite{Ball:2017otu,Nocera:2014gqa} at the initial scale $Q_0 = 1.64$ GeV. The statistical uncertainty of the fit was obtained via the widely known bootstrap method.
We refer to Ref.\tcite{Bacchetta:2020vty} for details on the fitting procedure and quality.
We stress that since our tree-level approximation does not take into account the gauge link, our model $T$-even TMDs are process-independent.
Preliminary results for spectator-model $T$-odd TMDs at twist-2 and their dependence on the gauge link can be found in Refs.\tcite{Bacchetta:2021lvw,Bacchetta:2021twk,Bacchetta:2022esb}.

Pursuing the goal of shedding light on the full 3D dynamics of gluons inside the proton, we consider the following densities which describe the 2D $\boldsymbol{p}_T$-distribution of gluons for different combinations of their polarization and the proton spin. For an unpolarized proton, we identify the unpolarized density 
\begin{equation}
 x \rho (x, p_x, p_y) = x f_1^g (x, \boldsymbol{p}_T^2) 
\label{eq:rho_unpol}
\end{equation}
as the probability density of finding unpolarized gluons at given $x$ and $\boldsymbol{p}_T$, while the Boer--Mulder distribution 
\begin{equation}
 x \rho^{\leftrightarrow} (x, p_x, p_y) = \frac{1}{2} \bigg[ x f_1^g (x, \boldsymbol{p}_T^2) + \frac{p_x^2 - p_y^2}{2 M^2} \, x h_1^{\perp g} (x, \boldsymbol{p}_T^2) \bigg]
\label{eq:rho_T}
\end{equation}
represents the probability density of finding linearly-polarized gluons in the transverse plane at $x$ and $\boldsymbol{p}_T$.

\begin{figure}[tb]
 \centering
 \includegraphics[scale=0.22,clip]{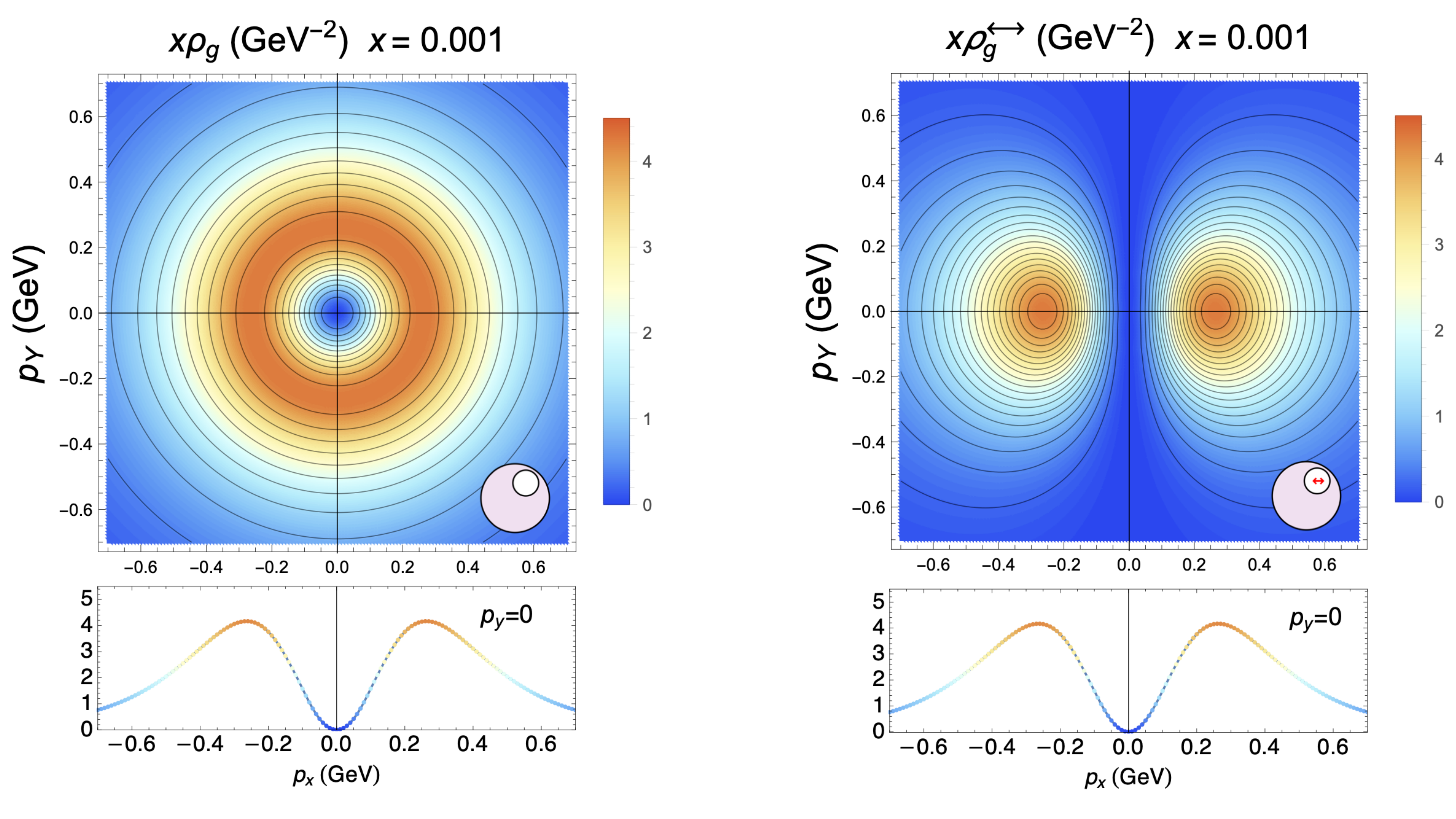}

 \caption{
 3D tomographic imaging of the proton  unpolarized (left) and Boer--Mulders (right) gluon TMD densities as functions of the gluon transverse momentum, for $x = 10^{-3}$ and at the initial energy scale, $Q_0 = 1.64$ GeV. 1D ancillary panels below main contour plots show the density at $p_y = 0$.
 Figures from~\cite{Bacchetta:2020vty}.
 }
 \label{fig:gluon_TMDs}
\end{figure}

Contour plots in Fig.~\ref{fig:gluon_TMDs} show $\boldsymbol{p}_T$-shape of the $\rho$-distributions in Eqs.~\eqref{eq:rho_unpol} and~\eqref{eq:rho_T}, respectively, obtained at $Q_0 = 1.64$ GeV and $x=10^{-3}$ for an unpolarized proton virtually moving towards the reader. The color code quantifies the size of the oscillation of each distribution along the $p_x$ and $p_y$ directions. To better catch these oscillations, ancillary 1D plots representing the corresponding density at $p_y = 0$ are shown below each contour plot. As expected, the density of Eq.~\eqref{eq:rho_unpol} exhibits a cylindrical symmetry around the direction of motion of the proton pointing towards the reader. Since the nucleon is unpolarized but the gluons are linearly polarized along the $p_x$ direction, the Boer--Mulders $\rho$-density in Eq.~\eqref{eq:rho_T} shows a dipolar structure. The departure from the cylindrical symmetry is emphasized at small-$x$, because the Boer--Mulders function is particularly large. 
From the analytic point of view, one has that the ratio between $f_1^g$ and $h_1^{\perp g}$ TMDs goes to a constant in the asymptotic small-$x$ limit, $x \to 0^+$.
This is in line with the prediction coming from the linear BFKL evolution, namely that at low-$x$ the "number" of unpolarized gluons is equal to the number of linearly-polarized ones, up to higher-twist effects (see, \emph{e.g.}, Refs.\tcite{Dominguez:2011br,Marquet:2016cgx,Taels:2017shj,Marquet:2017xwy,Petreska:2018cbf}). 
Thus, a connection point between our model gluon TMDs and the high-energy dynamics has been discovered.

\vspace{-0.20cm}
\section{Unintegrated gluon distribution}
\label{sec:UGD}
\vspace{-0.20cm}

The BFKL approach\tcite{Fadin:1975cb,Kuraev:1977fs,Balitsky:1978ic} affords us a factorized formula for scattering amplitudes (and thence, in inclusive reactions, for cross sections) given as a convolution of the universal BFKL Green's function and two process-dependent impact factors describing the transition from each initial-state particle to the corresponding detected object.

The connection between the UGD and gluon TMDs is still largely uncharted. From a formal perspective, the TMD formalism relies on parton correlators and thus on Wilson lines, whereas the BFKL approach ``speaks" the language of Reggeized gluons. From a phenomenological viewpoint, TMD factorization is expected to hold at low transverse momenta, whereas the BFKL resummation requires large transverse-momentum emissions.
A first connection between the UGD and the unpolarized and linearly polarized gluon TMDs, $f^g_1$ and $h^{\perp g}_1$,
was investigated in Refs.\tcite{Dominguez:2011wm,Hentschinski:2021lsh,Nefedov:2021vvy}.

The first class of processes that serves as probe channels of the BFKL dynamics is represented by the inclusive \emph{semi-hard} emission\tcite{Gribov:1983ivg} of two particles with high transverse momenta and well separated in rapidity (see Refs.\tcite{Celiberto:2017ius,Celiberto:2020wpk} for an overview of recent applications). Here the established factorization is \emph{hybrid}. Indeed the pure high-energy factorization is supplemented by collinear densities which enter expressions of impact factors. 

In the last thirty years several phenomenological studies have been proposed for different semi-hard final states. 
An incomplete list includes: the inclusive hadroproduction of two jets featuring large transverse momenta and well separated in rapidity (Mueller--Navelet channel\tcite{Mueller:1986ey}), for which several phenomenological studies have appeared so far~(see, \emph{e.g.},~Refs.\tcite{Colferai:2010wu,Caporale:2012ih,Ducloue:2013hia,Ducloue:2013bva,Caporale:2013uva,Caporale:2014gpa,Colferai:2015zfa,Caporale:2015uva,Ducloue:2015jba,Celiberto:2015yba,Celiberto:2015mpa,Celiberto:2016ygs,Celiberto:2016vva,Caporale:2018qnm}), the inclusive detection of two rapidity-separated light-flavored bound states\tcite{Celiberto:2016hae,Celiberto:2016zgb,Celiberto:2017ptm,Celiberto:2017uae,Celiberto:2017ydk}, three- and four-jet hadroproduction\tcite{Caporale:2015vya,Caporale:2015int,Caporale:2016soq,Caporale:2016vxt,Caporale:2016xku,Celiberto:2016vhn,Caporale:2016djm,Caporale:2016lnh,Caporale:2016zkc}, $J/\Psi$-plus-jet\tcite{Boussarie:2017oae,Boussarie:2017xdy}, hadron-plus-jet\tcite{Bolognino:2019cac,Bolognino:2019yqj,Bolognino:2019cac}, Higgs-plus-jet\tcite{Celiberto:2020tmb,Celiberto:2021fjf,Celiberto:2021tky,Celiberto:2021txb,Celiberto:2020rxb,Celiberto:2021xpm}, heavy-light dijet system\tcite{Bolognino:2021mrc,Bolognino:2021hxx}, heavy-flavor\tcite{Celiberto:2017nyx,Bolognino:2019yls,Bolognino:2019ccd,Celiberto:2021dzy,Celiberto:2021fdp,Bolognino:2022wgl}, and forward Drell–Yan dilepton production with a possible backward-jet detection~\cite{Golec-Biernat:2018kem}. 
This permitted us to define BFKL-sensitive observables as well as to disengage the BFKL dynamics from collinear contaminations\tcite{Celiberto:2015yba,Celiberto:2015mpa,Celiberto:2020wpk}.

\begin{figure}[t]
\centering
\includegraphics[width=0.47\textwidth]{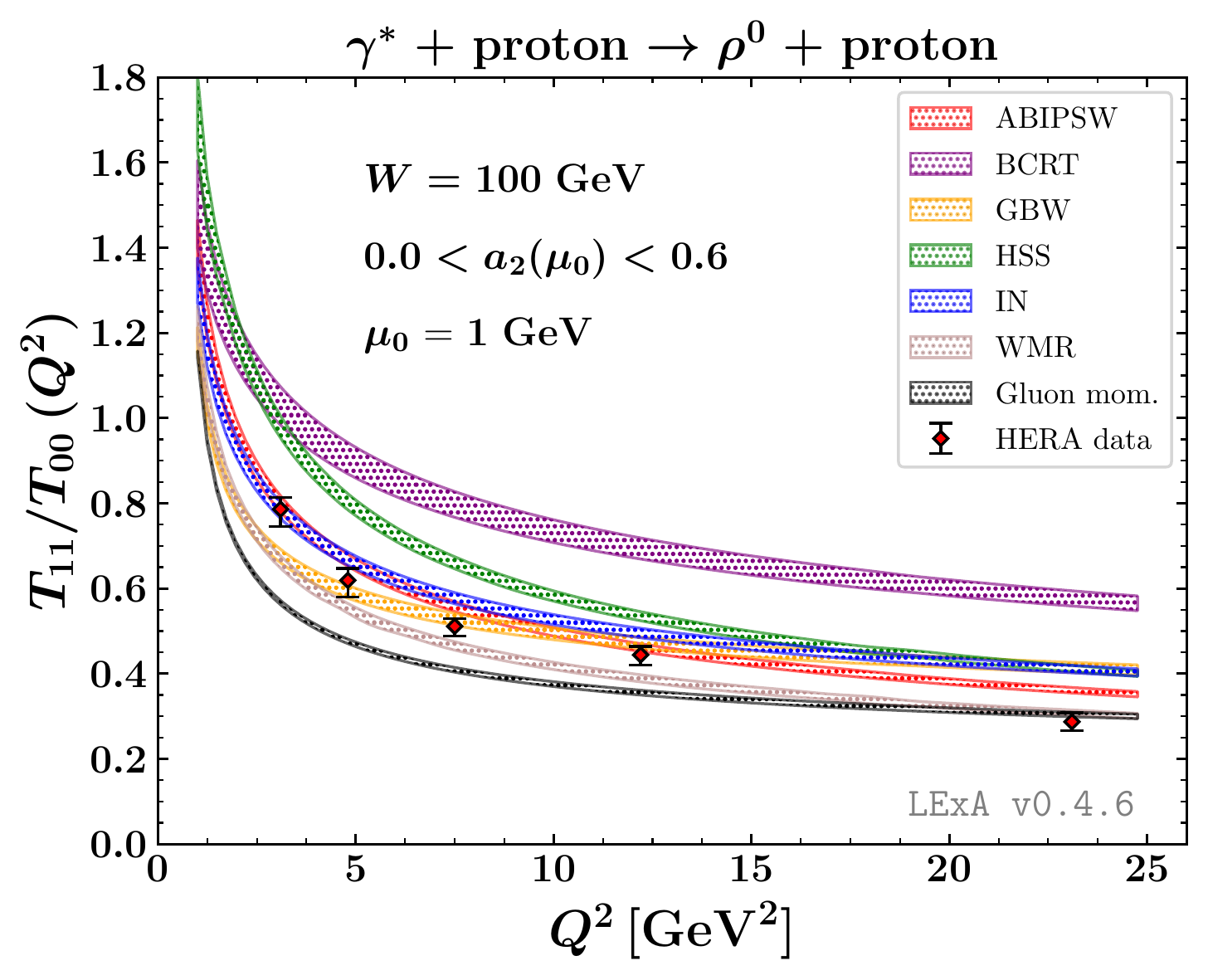} \hspace{0.5cm}
\includegraphics[width=0.47\textwidth]{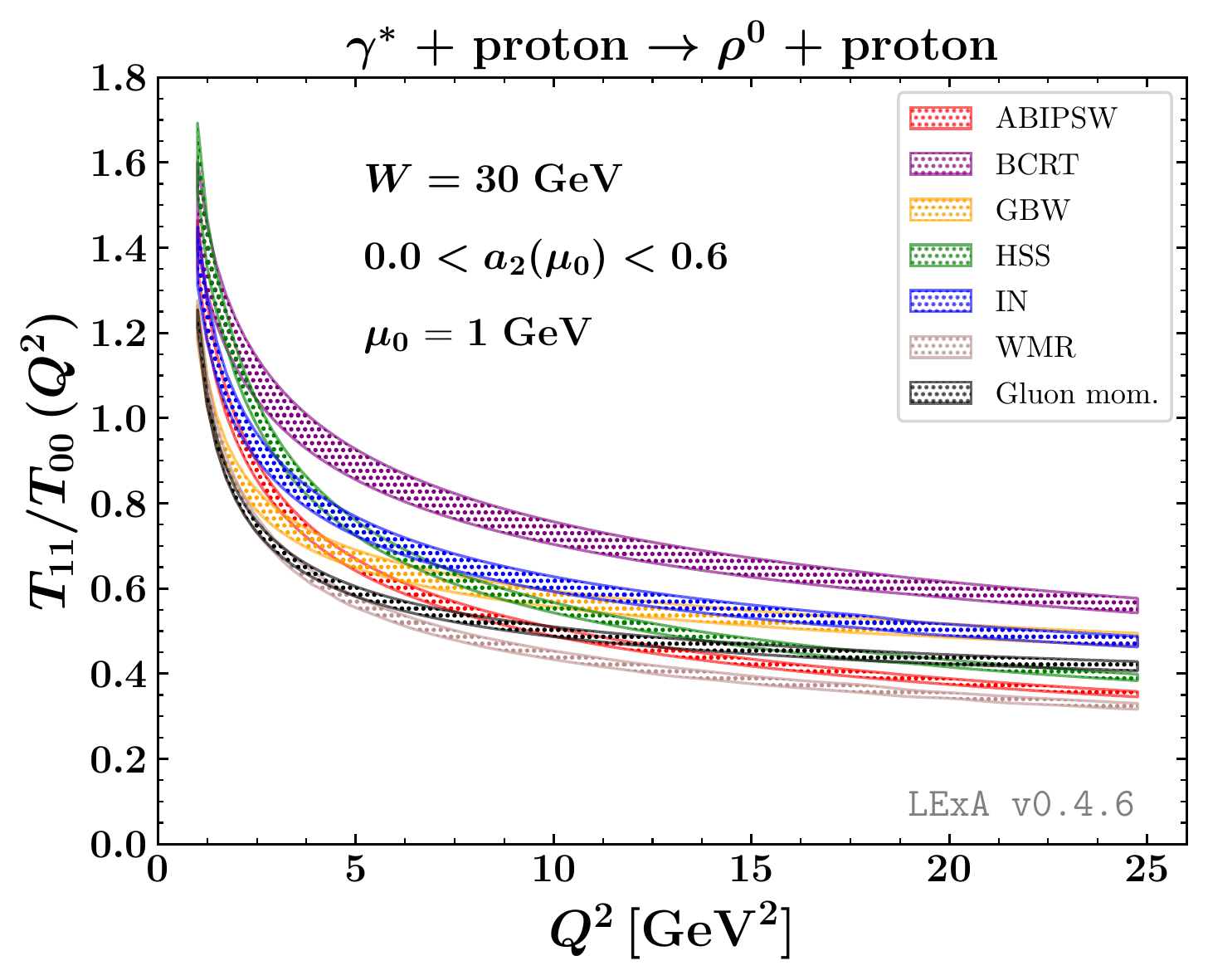}
\caption{$Q^2$-dependence of the polarized $T_{11}/T_{00}$ ratio, for all the considered UGD models, at $100$ GeV HERA (left) and at $30$ GeV EIC (right). 
Uncertainty bands represent the effect of varying $a_2(\mu_0 = 1\,$\rm GeV$)$ between $0.0$ and $0.6$.
Numerical results were obtained through the \emph{Leptonic-Exclusive-Amplitudes} ({\tt LExA}) super-module of the {\tt JETHAD} interface\tcite{Celiberto:2020wpk}. Figures from~\cite{Bolognino:2021niq}.}
\label{fig:rho}
\end{figure}

The second kind of high-energy probe channels is represented by single emissions of forward particles. Here we access the proton content through the UGD, whose evolution at small-$x$ is controlled by the BFKL Green's function. Being a non-perturbative object, the UGD in not well known and several phenomenological models for it have been built so far. The UGD has been probed via the inclusive deep inelastic scattering\tcite{Hentschinski:2012kr,Hentschinski:2013id}, the exclusive electro- or photo-production of vector mesons at HERA\tcite{Anikin:2009bf,Anikin:2011sa,Besse:2013muy,Bolognino:2018rhb,Bolognino:2018mlw,Bolognino:2019bko,Bolognino:2019pba,Celiberto:2019slj,Bautista:2016xnp,Garcia:2019tne,Hentschinski:2020yfm} and the EIC\tcite{Bolognino:2021niq,Bolognino:2021gjm,Bolognino:2022uty}, the single inclusive heavy-quark emission at the LHC\tcite{Chachamis:2015ona}, and the forward Drell--Yan production at LHCb\tcite{Motyka:2014lya,Brzeminski:2016lwh,Motyka:2016lta,Celiberto:2018muu}.

We consider the exclusive $\rho$-meson production in lepton-proton collisions via the subprocess
\begin{equation}
\label{eq:subprocess}
 \gamma^*_{\lambda_i} (Q^2) \, p \; \to \; \rho_{\lambda_f} p \;,
\end{equation}
where a photon with virtuality $Q^2$ and polarization $\lambda_i$ interacts with the proton and a $\rho$-meson with polarization $\lambda_f$ is produced. The two polarization states $\lambda_{i,f}$ can assume values $0$ (longitudinal) and $\pm 1$ (transverse).
Since a strict semi-hard scale ordering holds, $W^2 \gg Q^2 \gg \Lambda^2_{\rm QCD}$ ($W$ is the subprocess center-of-mass energy), one enters the small-$x$ regime given by $x = Q^2/W^2$. The BFKL approach provides us with a high-energy factorized formula for polarized amplitudes
\begin{equation}
\label{eq:ampltude}
 {\cal T}_{\lambda_i \lambda_f}(W^2, Q^2) = \frac{i W^2}{(2 \pi)^2} \int \frac{\drv^2 \boldsymbol{p}_T}{(\boldsymbol{p}_T^2)^2} \; \Phi^{\gamma^*_{\lambda_i} \to \rho_{\lambda_f}}(\boldsymbol{p}_T^2, Q^2) \, f_g^{\rm BFKL} (x, \boldsymbol{p}_T^2, Q^2) \;,
\end{equation}
with $\Phi^{\gamma^*_{\lambda_i} \to \rho_{\lambda_f}}(q^2, Q^2)$ being the impact factor that describes the $\gamma^* \to \rho$ transition and encodes collinear distribution amplitudes (DAs, for further details see Section~2 of Ref.\tcite{Bolognino:2021niq}), and $f_g^{\rm BFKL}$ is the BFKL UGD. We consider in our study the seven UGD models given in Section~3 of Ref.\tcite{Bolognino:2021niq}.

In Fig.\tref{fig:rho} we show the $Q^2$-dependence of ${\cal T}_{11} / {\cal T}_{00}$. We compare our predictions with HERA data\tcite{Aaron:2009xp} at $W = 100$ GeV (left panel), and we present new results for the EIC at the reference energy of $W = 30$ GeV (right panel). We use the twist-2 (twist-3) DAs for the longitudinal (transverse) case, and we gauge the impact of the collinear evolution of the DAs via a variation of the non-perturbative parameter $a_2(\mu_0 = 1\,$\rm GeV$)$ in the range 0.0 to 0.6.

We note that our predictions are spread over a large range and none of the UGD models is in agreement with HERA data over the whole $Q^2$-window, the ABIPSW, IN and GBW ones better catching the intermediate-$Q^2$ range. Results at EIC energies show a reduction of the distance between models, together with a hierarchy inversion for some regions of $Q^2$. This provides us with a clear evidence that the ${\cal T}_{11} / {\cal T}_{00}$ helicity ratio act as a discriminator for the UGD.

\vspace{-0.20cm}
\section{Future perspectives}
\label{sec:conclusions}
\vspace{-0.20cm}

We reported progresses on the study of the proton structure at small-$x$ via two distinct kinds of gluon distributions: the (un)polarized gluon TMD functions and the BFKL UGD.
All the presented results are relevant to explore the proton content at small-$x$, where the intrinsic motion of the constituent gluons plays an important role in the description of observables sensitive to different combinations of the hadron and the parton polarization states.
Here, a key ingredient to get a consistent description of the proton structure is interplay between genuine TMD and high-energy effects.
We believe that a path towards the first extraction of the small-$x$ improved gluon distributions from a global fit on data coming from new-generation colliding facility, such the EIC\tcite{Accardi:2012qut,AbdulKhalek:2021gbh}, the HL-LHC\tcite{Chapon:2020heu}, the FPF\tcite{Anchordoqui:2021ghd}, and NICA-SPD\tcite{Arbuzov:2020cqg} has been traced.

\vspace{-0.20cm}
\section*{Acknowledgements}
\vspace{-0.20cm}

We thank Alessandro Bacchetta, Andr\`ee Dafne Bolognino, Dmitry Yu. Ivanov, Alessandro Papa, Marco Radici, Wolfgang Sch\"afer, Antoni Szczurek, and Pieter Taels for collaboration.


\vspace{-0.20cm}
\bibliographystyle{apsrev}
\bibliography{references}

\end{document}